\documentclass{ws-procs9x6}

\begin{document}
\title{MSSM: source for generating perturbations  }

\author{A. Mazumdar}

\address{McGill University, 3600 University Rue, Montr\'eal,
QC, H3A 2T8, Canada}


\maketitle

\abstracts{In this talk I describe how MSSM flat directions 
act as a source for generating all matter and scale invariant
adiabatic density perturbations with a spectral index slightly
depending upon the choice of a particular flat direction.}

\section{Introducing MSSM flat directions}

Supersymmetry (SUSY) is by far the best candidate beyond the
electroweak Standard Model. It explains the hierarchy in scales,
resolves the Higgs stability, gives rise to gauge unification, and
foremost it is a building block for superstring
theories~\cite{Nilles}. In this talk I will concentrate upon its
cosmological consequences~\cite{Enqvist}. Cosmology is a booming area
where recent satellite based experiments have given a new precession
measurements of various cosmological parameters~\cite{WMAP}.

The highlight of the talk is to show that Minimal Supersymmetric
Standard Model (MSSM) can act as a source for all matter and
primordial density perturbations. The is because MSSM possesses F-and
D-flat directions (made up of squarks and sleptons). For a generic
SUSY model, with chiral superfields $X_{i}$, it is possible to find
out the directions, where $N=1$ SUSY potential, $V=V_{F}+V_{D}$,
vanishes identically, by solving simultaneously
\begin{equation}
\label{fflatdflat}
D^{a}\equiv X^{\dagger}T^{a}X=0\,, \quad \quad
F_{X_{i}}\equiv \frac{\partial W}{\partial X_{i}}=0\,.
\end{equation}
Field configurations obeying Eq.~(\ref{fflatdflat}) are called
respectively D-flat and F-flat. These directions are parameterized by
gauge invariant monomials of the chiral superfields. A powerful tool
for finding the flat directions has been developed in Ref.~\cite{Buccella},
where the correspondence between a gauge invariant monomial and flat
directions has been studied. Its generalization to gauge invariant
polynomials can be found in Ref.~\cite{Jokinen}. These flat directions act
as a cosmic condensate during inflation. There are various
cosmological consequences which I will {\it not} discuss here,
interested readers are referred to Ref.~\cite{Enqvist}.

\section{Inflationary paradigm}

Inflation is the most favored paradigm for the early Universe. Besides
making the Universe flat, homogeneous and isotropic, it is the only
causal mechanism which stretches the fluctuations (below the Planckian
energy density) outside the horizon. After inflation these modes
re-enter and act as seeds for the structure formation. In spite of all
these successes the identity of inflaton is not known, nor its
potential. It is often regarded as an absolute gauge singlet.
Inflationary sector is almost like a dark energy sector, which is
hard to pin down without a knowledge of full quantum nature of
gravity. Very recently it has been possible to show that gauge
invariant flat directions can give rise to inflation in $SU(n),~SO(n)$
theories~\cite{Asko}.

\section{Forming a cosmic condensate}

During inflation squarks and sleptons are free to fluctuate along the
flat directions and form scalar condensates. Because inflation
smoothes out all gradients, only the homogeneous condensate mode
survives. However, like any massless scalar field, the condensate is
subject to inflaton-induced zero point fluctuations which impart a
small, and in inflation models a calculable, spectrum of perturbations
on the condensate. During inflation the dynamical evolution of these
condensates is frozen, but after inflation their dynamics play
important role~\cite{Enqvist}.

In addition to the usual softly SUSY breaking, the non-zero energy
density of the early Universe also breaks SUSY, in particular during
inflation when the Hubble expansion dominates over any low energy SUSY
breaking scales \cite{Gherghetta}. Flatness can also be spoiled by
higher-order non-renormalizable terms. A generic potential during
inflation is a sum of flat direction potential $V(\phi)$ and the 
inflaton potential $V(I)$, where
\begin{equation}
\label{pot}
V(\phi)\sim (1/2)m_{3/2}^2|\phi|^2+
C_{H}H^2|\phi|^2+\lambda^2|\phi|^{2n-2}/ M_{p}^{2n-6}\,,
\end{equation}
where $H$ is the expansion rate, and $n=4,..,9$ (I neglected the
A-terms, for a full flat direction potential, see~\cite{Enqvist}),
$m_{3/2}\sim O(1)$~TeV (m-SUGRA), $\lambda\sim O(1)$, and $C_{H}\sim
O(1)$ for a minimal K\"ahler structure. For no-scale SUGRA, one-loop
correction gives~$C_{H}\sim 0.01$~\cite{Gaillard}. For our discussion
we assume the Hubble induced mass term is vanishing,
e.g. $C_{H}\approx 0$, the situation may arise in no-scale SUGRA
model, or inflation driven by D-term potential, or a non-minimal
K\"ahler structure. For $C_{H}\neq 0$, the flat direction simply rolls
down the potential during inflation, leaving rather uninteresting
consequences.

Not all F- and D-flat directions can act as a condensate, only one out of
nearly 300 MSSM flat directions can obtain a large vev, because not
all of them are independent, for
e.g. $LH_{u},~LLe,~H_{u}H_{d}$ are not independent. 

\section{Density perturbations}
The initial density perturbations generated by the MSSM flat
directions is in the form of isocurvature perturbations, because there
are at least two fields dynamically evolving, inflaton and the flat
direction. For multi fields the total curvature perturbations,
$\zeta$, outside the horizon is not constant, because there is a
pressure difference between the fields which is evolving. This
isocurvature perturbations has to be converted into the adiabatic
perturbations. This happens after inflation, $H$ drops fast,
when $m_{3/2}\sim H(t)$, then the flat direction starts
oscillating and decaying into MSSM degrees of freedom~\cite{Mari}.
However it is important that its energy density must dominate while
decaying. This puts severe constraint on MSSM flat direction, only
candidates are $LLddd$ (lifted by $n=7$ non-renormalizable operator,
$H_{u}LLLddd$), and $QuQuQue$ (lifted by $n=9$ operator,
$QuQuQuH_{d}ee$)~\cite{Enqvist1}.

The curvature perturbations for the flat direction is given by 
$P^{1/2}_{\zeta}\approx rH_{\ast}/\pi\phi_{\ast}$~\cite{Lyth}, where 
$r={3 \rho_\phi}/({4 \rho_\gamma + 3 \rho_\phi})$, where
$\rho_\gamma$ is the energy in the radiation bath from inflaton
decay. Non-Gaussianity of the produced perturbation requires the
curvaton to contribute more than 1\% to the energy density of the
universe at the time of decay, that is $r_{\rm dec} > 0.01$~\cite{Lyth}.
During inflation only the non-renormalizable term dominates, 
Eq.~(\ref{pot}), which gives an interesting relationship
\begin{equation}
H_{\ast}\sim \beta^{(1/n-3)}\delta^{(n-2/n-3)}M_{p}\,,~~~
\phi_{\ast}\sim \beta^{(1/n-3)}\delta^{(1/n-3)}M_{p}\,,~~\delta \sim 
(H_{\ast}/\phi_{\ast})\,.\nonumber \\
\end{equation}
where $\ast$ denotes the epoch during inflation when modes are leaving
the horizon, and the amplitude of the fluctuations, $\delta
=\delta\phi/\phi_{\ast} \sim 10^{-5}$.  The value of $\beta\ll 1$
($\beta$ depends on the superpotential which lifts the flat direction)
arises from $V^{\prime\prime}(\phi)\sim \beta^2H_{\ast}^2\ll
H_{\ast}^2$.  The spectral index of the perturbations is then given
by~\cite{Lyth}
\begin{eqnarray}
n_s-1\sim 2{\dot{H_*}}/{H_*^2}+
({2}/{3})({V''(\phi_*)}/{H_*^2})\sim (2/3)\beta^2\,,
\end{eqnarray}
where we assumed that the Hubble expansion is nearly a constant,
$\dot H/H^2\approx 0$. Note that for $\beta<0.1$, the spectral index
is close to the observable limit $n_s=0.99 \pm 0.04$~\cite{WMAP}.

In an another paradigm shift, it is even conceivable to dump the
inflaton energy density outside our world, e.g., in an anti-de-Sitter
(adS) background geometry, where the observable world is a
$3$-dimensional hypersurface (realizable in string theory via stack of
$D3$, $D7$ branes), the inflaton energy density can be dumped near the
adS throat~\cite{Kasuya,lorenzana}. For the observable world the
dumped energy will be red-shifted because of the non-trivial warped
background.  This miniscule energy (depending on the size of the
overall manifold, string coupling and string scale) could even answer
our quest for the dark energy. In such a set up, after inflation, the
only energy density is left behind is the condensate energy. In this
case the best candidate for the flat direction is the MSSM Higgses
$H_{u}H_{d}$~\cite{Kasuya}. In both the cases, discussed above, the
reheat temperature, $T_{rh}\leq 10^{9}$~GeV~\cite{Enqvist1,Kasuya}, is
below the gravitino overproduction limit~\cite{Gravi}.

Yet another possibility arises when the inflaton and the MSSM flat
directions are not completely independent, they can have
non-renormalizable coupling. In which case the inflaton coupling to
the MSSM obtains fluctuating inflaton coupling which gives rise to
the spatial fluctuations in the decay rate and the reheating
temperature~\cite{Dvali,Marieke}.

I wish to thank all my collaborators.                  
A.M is a CITA-National fellow.                  


\end{document}